\documentclass[11pt]{article}
\usepackage{amsmath}
\usepackage{amsfonts}
\usepackage{graphics}
\usepackage{graphicx}
\usepackage{ragged2e}
\usepackage{float}
\usepackage{latexsym}
\usepackage{amsthm}
\usepackage{cite}
\usepackage{amssymb}
\usecounter{enumi}
\newtheorem{Lemma}{Lemma}[section]
\newtheorem{Theorem}[Lemma]{Theorem}
\newtheorem{Example}[Lemma]{Example}
\newtheorem{Remark}[Lemma]{Remark}
\newtheorem{Corollary}[Lemma]{Corollary}
\newtheorem*{Proof}{Proof}
\newtheorem{Proposition}[Lemma]{Proposition}

\numberwithin{equation}{section}

\def\be{\begin{eqnarray}} \def\ee{\end{eqnarray}} \def\k{\kappa}
  \def\({\left(} \def\){\right)}
\def\bc{\begin{center}} \def\l{\label}
\def\ec{\end{center}}  

\def\bey{\begin{eqnarray*}}\def\eey{\end{eqnarray*}}
\begin{document}
\title{Systems of partial differential equations describing pseudospherical or spherical surfaces}

\author{ Mingyue Guo$^{1}$, Jing Kang$^{1,2,3}$, Zhenhua Shi$^{1,2}$
\vspace{4mm}\\
$^{1}$\small{School of Mathematics, Northwest University, Xi'an 710069,  P.R. China}\\
$^{2}$\small {Center for Nonlinear Studies, Northwest University, Xi'an 710069, P.R. China}\\
$^{3}$\small {Shaanxi Key Laboratory of Mathematical Theory and Computation of Fluid Mechanics,} \\
 \small {Northwest University, Xi'an 710069, P.R. China}}

\date{}
\maketitle
\begin{minipage}{140mm}

\begin{abstract}
In this paper, we study systems of nonlinear partial differential equations which describe surfaces of constant curvature. From the flatness condition of connection 1-forms, we present a classification of systems of Camassa-Holm-type equations of the form
\begin{equation*}
\left\{
\begin{aligned}
u_{t} - u_{xxt} &= F(x, t, u, u_{x}, \dots, \partial ^{m} u/\partial _x^{m}, v, v_{x}, \dots, \partial ^{n} v/\partial _x^{n}), \\
v_{t} - v_{xxt} &= G(x, t, u, u_{x}, \dots, \partial ^{m} u/\partial _x^{m}, v, v_{x}, \dots, \partial ^{n} v/\partial _x^{n}),
\end{aligned}
\right.
\end{equation*}
with $m,n\geq2$, for $F$ and $G$ smooth functions, describing pseudospherical or spherical surfaces. We also establish classification results for a special type of third-order system. Applications of the results provide new examples of such systems, such as the Song-Qu-Qiao system, the two-component Camassa-Holm system with cubic nonlinearity, and the modified Camass-Holm-type system. Moreover, we construct the nonlocal symmetry and a non-trivial solutions for the two-component Camassa-Holm system with cubic nonlinearity from the gradients of spectral parameters.
\end{abstract}

\end{minipage}

\section{Introduction}

\indent \indent The concept of differential equations which describe pseudospherical surfaces was first introduced by Chern and Tenenblat \cite{chern1986}, building upon an earlier observation by Sasaki's observation \cite{sasaki1979}. They noted that all $1+1$-dimensional soliton equations solvable by the AKNS $2\times2$ inverse scattering method describe pseudospherical surface. A classical example is the sine-Gordon (SG) equation, originally discovered by Bour \cite{bour1862}. Subsequent works have significantly deepened the understanding of such equations, as displayed in \cite{jorge1987,rabelo1989,rabelo1992,kamran1995,reyes1998,reyes2006,reyes2011,neto2010,ferraioli2014,ferraioli2016,ferraioli2020}.

The study of differential equations describing pseudospherical or spherical surfaces holds considerable theoretical and practical importance in mathematics and physics \cite{cavalcante1988}. Such equations can be analyzed using geometric methods, especially for their integrability conditions, which are linked to $\mathfrak{sl}(2, \mathbb{R})$-valued or $\mathfrak{su}(2)$-valued linear problems, allowing the construction of solutions via the inverse scattering method \cite{ablowitz1974}. Moreover, generic solutions for these equations provide metrics on non-empty open subsets of $\mathbb{R}^2$ with constant Gaussian curvature $K=-1$ or $K=1$.

In 2002, Ding and Tenenblat \cite{ding2002} extended the notion of differential equations describing pseudospherical or spherical surfaces to differential systems that describe pseudospherical or spherical surfaces, providing a general characterization of evolution systems which describe pseudospherical or spherical surfaces. Specifically, they classified all differential systems of the form
\begin{equation}\label{1.1}
\left\{
\begin{aligned}
u_{t}  &= -v_{xx}+H_{11}(u,v)u_x+H_{12}(u,v)v_x+H_{13}(u,v),\\
v_{t}  &= u_{xx}+H_{21}(u,v)u_x+H_{22}(u,v)v_x+H_{23}(u,v),
\end{aligned}
\right.
\end{equation}
which describe $\eta$-pseudospherical or $\eta$-spherical surfaces. As applications, several significant equations were derived, including the nonlinear Schr\"odinger (NLS) equation, the Heisenberg Ferromagnet (HF) model and the Landau-Lifschitz equation.

In 2022, Kelmer and Tenenblat \cite{kelmer2022} established classification results for systems of partial differential equations of the form
\begin{equation}\label{1.2}
\left\{
\begin{aligned}
u_{t}  &= F(u,u_x,v,v_x),\\
v_{t}  &= G(u,u_x,v,v_x),
\end{aligned}
\right.
\end{equation}
describing pseudospherical or spherical surfaces. These systems contain generalizations of a Pohlmeyer-Lund-Regge type system and the Konno-Oono coupled dispersionless system. More recently, the same authors \cite{kelmer2025} characterized systems of partial differential equations describing pseudospherical or spherical surfaces of the form
\begin{equation}\label{1.3}
\left\{
\begin{aligned}
u_{xt}  &= F(u, u_{x}, \dots, \partial ^{n} u/\partial _x^{n}, v, v_{x}, \dots, \partial ^{m} v/\partial _x^{m}),\\
v_{xt}  &= G(u, u_{x}, \dots, \partial ^{n} u/\partial _x^{n}, v, v_{x}, \dots, \partial ^{m} v/\partial _x^{m}),
\end{aligned}
\right.
\end{equation}
where $n,m\geq2$, and $F$, $G$ are smooth functions. Notable examples include the vector short-pulse and its generalizations.

Kelmer \cite{kelmer2024} further investigated systems of third-order evolution equations of the form
\begin{equation}\label{1.4}
\left\{
\begin{aligned}
u_{t}  &= F(x,t,u, u_{x}, u_{xx},u_{xxx}, v, v_{x}, v_{xx},v_{xxx}),\\
v_{t}  &= G(x,t,u, u_{x}, u_{xx},u_{xxx}, v, v_{x}, v_{xx},v_{xxx}),
\end{aligned}
\right.
\end{equation}
describing pseudospherical or spherical surfaces. Applications of these results yield new families of such systems, including the coupled Korteweg-de Vries (KdV) system, the mKdV-type systems and the third-order NLS-type systems.

Over the past three decades, the Camassa-Holm (CH)-type equations have been among the most widely studied integrable systems \cite{novikov2009}. Although significant progress has been made in understanding CH-type equations which describe pseudospherical or spherical surfaces \cite{reyes2002,silva2015,ferraioli2024,guo2025}, much less is known regarding systems of CH-type equations. On the other hand, it has been established that the geometric approach, used to determine whether a given differential equation describes pseudospherical or spherical surfaces, can be extended to systems of differential equations. It is therefore of considerable interest to investigate whether systems of CH-type differential equations can describe pseudospherical or spherical surfaces, and whether their integrability admits a geometric interpretation via an $\mathfrak{sl}(2, \mathbb{R})$-valued linear problem. Inspired by these questions, this paper aims to classify systems of partial differential equations of the following form
\begin{equation}\label{1.5}
\left\{
\begin{aligned}
u_{t} - u_{xxt} &= F(x, t, u, u_{x}, \dots, \partial ^{m} u/\partial _x^{m}, v, v_{x}, \dots, \partial ^{n} v/\partial _x^{n}), \\
v_{t} - v_{xxt} &= G(x, t, u, u_{x}, \dots, \partial ^{m} u/\partial _x^{m}, v, v_{x}, \dots, \partial ^{n} v/\partial _x^{n}),
\end{aligned}
\right.
\end{equation}
which describe pseudospherical or spherical surfaces, with $m,n\geq2$, for $F$ and $G$ smooth functions. The classification problem corresponds to determine the system \eqref{1.5} admitting 1-forms
\begin{equation}\label{1.6}
\omega_i = f_{i1}\,dx + f_{i2}\,dt, \quad 1 \leq i \leq 3,
\end{equation}
where coefficient functions $\displaystyle f_{ij}=f_{ij}(x,t,u,u_{x}, \dots, \partial ^{m} u/\partial _x^{m}, v, v_{x}, \dots, \partial ^{n} v/\partial _x^{n})$ satisfy a particular system of equations.

In this paper, we show families of systems of partial differential equations contained in classification theorems. For examples,

\noindent (i) The Song-Qu-Qiao system \cite{song2011,tian2013,chang2016,qiao2015a}
\begin{equation}\label{1.7}
\left\{
\begin{aligned}
u_{t} - u_{xxt} &= [(u-u_{xx})(u_x v_x -uv+uv_x-u_xv)]_x, \\
v_{t} - v_{xxt} &= [(v-v_{xx})(u_x v_x -uv+uv_x-u_xv)]_x.
\end{aligned}
\right.
\end{equation}

\noindent (ii) The two-component CH system with cubic nonlinearity \cite{qiao2013,qiao2015b,xia2015a}
\begin{equation}\label{1.8}
\left\{
\begin{aligned}
u_{t} - u_{xxt} &= \frac{1}{2}[(u-u_{xx})(uv-u_x v_x)]_x-\frac{1}{2}(u-u_{xx})(u v_x -u_x v), \\
v_{t} - v_{xxt} &= \frac{1}{2}[(v-v_{xx})(uv-u_x v_x)]_x+\frac{1}{2}(v-v_{xx})(u v_x -u_x v).
\end{aligned}
\right.
\end{equation}

\noindent (iii) The following system \cite{xia2015b}
\begin{equation}\label{1.9}
\left\{
\begin{aligned}
u_{t} - u_{xxt} &= -\frac{1}{2}(u-u_{xx})(u-u_x )(v+v_x), \\
v_{t} - v_{xxt} &=\frac{1}{2}(v-v_{xx})(u-u_x)(v+v_x).
\end{aligned}
\right.
\end{equation}

\noindent (iv) The modified CH-type system
\begin{equation}\label{1.10}
\left\{
\begin{aligned}
u_{t} - u_{xxt} &=-\left[\left(\frac{1}{2}(u^2 +v^2-u_x^2-v_x^2)+(uv_x -u_x v)\right)(u-u_{xx})\right]_x-2u_x , \\
v_{t} - v_{xxt} &= -\left[\left(\frac{1}{2}(u^2 +v^2-u_x^2-v_x^2)+(uv_x -u_x v)\right)(v-v_{xx})\right]_x-2v_x.
\end{aligned}
\right.
\end{equation}

In the study of integrable CH-type equations, nonlocal symmetries mean infinitesimal symmetries non-trivially depending on potentials or pseudo-potentials \cite{shchik1989,galas1992,guthrie1994}. Nonlocal symmetries with pseudo-potentials were calculated for the CH equation \cite{reyes2002}, the modified CH equation \cite{reyes2007,bies2012}, the 2-component CH equation, 2-component generalization of the modified CH equation \cite{tian2024}  and etc.. Inspired by these works, we want to explore nonlocal symmetries of the two-component CH system with cubic nonlinearity \eqref{1.8}, which is an isospectral flow of a linear spectral problem and admit bi-Hamiltonian structure. Thus, its spectral parameter is treated as a conserved quantity. For any Hamiltonian system, the Hamiltonian operator would map gradients of conserved quantities into its infinitesimal symmetry. Therefore, applying Hamiltonian operator to gradients of spectral parameter would produce nonlocal infinitesimal symmetries depending on eigenfunctions of linear spectral problems. By this approach, we calculate the nonlocal symmetries for the system \eqref{1.8} and prolong to an enlarged system (consisting of this system, its linear problem and equations defining an auxiliary pseudo-potential), from which generate a finite symmetry transformation for the enlarged system. As further applications, we derive a non-trivial solution for the system \eqref{1.8}.

The remainder of this paper is organized as follows. In Section 2, we give a brief review in some basic results concerning the systems of partial differential equations related to pseudospherical or spherical surfaces, in terms of linear problems associated to $\mathfrak{sl}(2, \mathbb{R})$-valued connection 1-form $ \omega_i $. The main results for classification will be shown in Section 3, Theorem 3.4-3.7, with explicit examples such as the Song-Qu-Qiao system, the two-component CH system with cubic nonlinearity and the modified CH-type system. The nonlocal symmetry and a non-trivial solution for the two-component Camassa-Holm system with cubic nonlinearity are presented in Section 4. And the last section is left for conclusion and discussion.

\section{Preliminaries}

\indent \indent In 1979, Wadati et al. \cite{wadati1979} studied the inverse scattering problem
\begin{equation}\label{2.1}
V_x=XV,\quad X=\left(
               \begin{array}{cc}
                 F(\eta) & H(\eta)q(x,t) \\
                 H(\eta)r(x,t) & -F(\eta) \\
               \end{array}
             \right),
\end{equation}
where $F(\eta)$ and $H(\eta)$ are functions of the spectral parameter $\eta$, and the time evolution of the eigenfunctions is governed by
\begin{equation}\label{2.2}
V_t=TV,\quad T=\left(
               \begin{array}{cc}
                 A(\eta,q,r) & B(\eta,q,r) \\
                 C(\eta,q,r) & -A(\eta,q,r) \\
               \end{array}
             \right),
\end{equation}
with $V = (v_1, v_2)^T$ and $v_i=v_i(x,t)$. By imposing the compatibility condition $V_{xt}=V_{tx}$ and assuming that the eigenvalues $\eta$ are time-invariant, we obtain zero curvature representation \cite{crampin1977}
\begin{equation}\label{2.3}
X_t - T_x + XT-TX = 0,
\end{equation}
which leads to the system for the functions $A(\eta,q,r)$, $B(\eta,q,r)$, $C(\eta,q,r)$
\begin{equation}\label{2.4}
\begin{aligned}
  A_x+H(rB-qC)&=0,  \\
  Hq_t-B_x-2FB-2HqA&=0, \\
  Hr_t-C_x+2FC+2HrA&=0.
\end{aligned}
\end{equation}

Using exterior calculus, the inverse scattering problem \eqref{2.1} and \eqref{2.2} is reformulated as a completely integrable linear system
\begin{equation}\label{2.5}
	dV = \Omega V,
\end{equation}
where $\Omega$ is a traceless $2\times 2$ matrix, given by the $\mathfrak{sl}(2, \mathbb{R})$-valued 1-form
\begin{equation}\label{2.6}
\Omega = \frac{1}{2} \begin{pmatrix} \omega_2 & \omega_1 - \omega_3 \\ \omega_1 + \omega_3 & -\omega_2 \end{pmatrix},
\end{equation}
with the associated 1-forms $\omega_i$ defined by
\begin{equation}\label{2.7}
\begin{aligned}
  \omega_1&=H(r+q)\,dx+(B+C)\,dt,  \\
  \omega_2&=2F \,dx+2A\,dt,\\
  \omega_3&=H(r-q)\,dx+(C-B)\,dt.
\end{aligned}
\end{equation}
The integrability condition for \eqref{2.5} is
\begin{equation}\label{2.8}
	d\,\Omega - \Omega \wedge \Omega = 0,
\end{equation}
which is equivalent to the following relations
\begin{equation}\label{2.9}
d\omega_1 = \omega_3 \wedge \omega_2, \quad d\omega_2 = \omega_1 \wedge \omega_3,\quad d\omega_3 = \omega_1 \wedge \omega_2.
\end{equation}
In solving \eqref{2.4} or \eqref{2.8} for the functions $F,H,A,B,C$, it is generally necessary to satisfy an additional partial differential equation.

Suppose $S$ is a two-dimensional Riemannian manifold endowed with a coframe $\{\omega_1,\omega_2\}$ dual to an orthogonal frame $\{e_1, e_2\}$. The metric on $S$ can be expressed as $g = \omega_1^2 + \omega_2^2$. The first two equations in \eqref{2.9} are the structure equations determining the connection form $\omega_3:=\omega_{12}$, while the last equation in \eqref{2.9}, known as the Gauss equation, implies that the Gaussian curvature of $S$ is -1, meaning $S$ is a pseudospherical surface.

A system of partial differential equation for scalar functions $u(x, t)$ and $v(x,t)$ is said to describe pseudospherical surfaces ($\delta = 1$) or spherical surfaces ($\delta = -1$) if there exist 1-forms $\omega_i = f_{i1} dx + f_{i2} dt$, $1 \leq i \leq 3$, with $\omega_1 \wedge \omega_2 \neq 0$, where the coefficient functions $f_{ij}$, $j = 1, 2$, depend on $x$, $t$, $u(x, t)$, $v(x,t)$ and its derivatives with respect to $x$ and $t$, such that the structure equations of a surface with Gaussian curvature $K = -1$ or $K = 1$, say
\begin{equation}\label{2.10}
	d\omega_1 = \omega_3 \wedge \omega_2, \quad d\omega_2 = \omega_1 \wedge \omega_3, \quad d\omega_3 = \delta \omega_1 \wedge \omega_2,
\end{equation}
are satisfied whenever ($u(x,t)$, $v(x,t)$) is a solution of the system.

\begin{Remark}
The $2\times 2$ matrix $\Omega$ is not unique for linear problem \eqref{2.5} and the integrability condition \eqref{2.8} is invariant under the gauge transformation \cite{sasaki1979}
\begin{equation}\label{2.11}
\Omega \rightarrow \Omega^{\prime}=dAA^{-1}+A\Omega A^{-1},
\end{equation}
with $A$ a $2\times2$ matrix satisfying $\det A=1$. In fact, choosing
\begin{equation}\label{2.12}
A=\frac{\sqrt{2}}{2} \begin{pmatrix}-i & 1 \\ 1 & -i\end{pmatrix},
\end{equation}
we have the $\mathfrak{su}(2)$-valued 1-form
\begin{equation}\label{2.13}
\Omega = \frac{1}{2} \begin{pmatrix} i\omega_3 & \omega_1 -i\omega_2 \\ \omega_1 + i\omega_2 & -i\omega_3 \end{pmatrix}.
\end{equation}
\end{Remark}

\begin{Remark}
Note that the local isomorphism between $SL(2,\mathbb{R})$ and $SO(2,1)$ provides the Lie algebra isomorphism $\mathfrak{so}(2,1)\cong\mathfrak{sl}(2,\mathbb{R})$, and the local isomorphism between $SO(3)$ and $SU(2)$ provides the Lie algebra isomorphism $\mathfrak{so}(3)\cong\mathfrak{su}(2)$ \cite{kelmer2022}. Hence, we find $\mathfrak{so}(2,1)$ (resp. $\mathfrak{so}(3)$)-valued 1-form
\begin{equation}\label{2.14}
\widetilde{\Omega}= \begin{pmatrix} 0 & \omega_1 & \omega_2 \\ \delta \omega_1 & 0 & \omega_3 \\ \delta \omega_2 & -\omega_3 & 0 \end{pmatrix},
\end{equation}
with $\delta = 1$ (resp. $\delta = -1$).
\end{Remark}

Currently, numerous systems of partial differential equations are known to describe pseudospherical or spherical surface.  A particularly representative example the defocusing $NLS^{-}$ equation \cite{ding2002}
\begin{equation}\label{2.15}
\begin{cases}
u_t+v_{xx}-2(u^2+v^2)v=0,\\
-v_t+u_{xx}-2(u^2+v^2)u=0,
\end{cases}
\end{equation}
with associated 1-forms
\begin{equation}\label{2.16}
\begin{aligned}
\omega_1&=2u\,dx-2(2\eta u+v_x)\,dt, \\
\omega_2&=-2v\,dx+2(2\eta v-u_x)\,dt, \\
\omega_3&=2\eta\,dx-2(2\eta^2+u^2+v^2)\,dt, \\
\end{aligned}
\end{equation}
where $\eta\in\mathbb{R}$ is a spectral parameter. Indeed, the system \eqref{2.15} is equivalent to the structure equations \eqref{2.10} with $\delta=1$.

On the other hand, a well-known example of a system of partial differential equation describing spherical surfaces is the focusing $NLS^{+}$ equation \cite{ding2002}
\begin{equation}\label{2.17}
\begin{cases}
u_t+v_{xx}+2(u^2+v^2)v=0,\\
-v_t+u_{xx}+2(u^2+v^2)u=0,
\end{cases}
\end{equation}
with associated 1-forms
\begin{equation}\label{2.18}
\begin{aligned}
\omega_1&=2v\,dx-2(2\eta v-u_x)\,dt, \\
\omega_2&=2\eta\,dx-2(2\eta^2-u^2-v^2)\,dt, \\
\omega_3&=-2u\,dx+2(2\eta u+v_x)\,dt, \\
\end{aligned}
\end{equation}
where $\eta\in\mathbb{R}$ is a spectral parameter. In fact, the system \eqref{2.17} is equivalent to the structure equations \eqref{2.10} with $\delta=-1$.

Hereafter, for the convenience of discussion, we adopt the notation,
\begin{equation}\label{2.19}
\begin{aligned}
u_1 &= \frac{\partial u}{\partial x}, & u_2 &= \frac{\partial^2 u}{\partial x^2}, & u_3 &= \frac{\partial^3 u}{\partial x^3}, & \dots,\quad u_m &= \frac{\partial^m u}{\partial x^m}, \\
v_1 &= \frac{\partial v}{\partial x}, & v_2 &= \frac{\partial^2 v}{\partial x^2}, & v_3 &= \frac{\partial^3 v}{\partial x^3}, & \dots,\quad v_n &= \frac{\partial^n v}{\partial x^n}.
\end{aligned}
\end{equation}

\section{Main results and Examples}
\indent \indent The present section will concentrate on the classification of the system of type
\begin{equation}\label{3.1}
\left\{
\begin{aligned}
u_{t} - u_{2,t} &= F(x, t, u, u_{1}, \dots, u_{m}, v, v_{1}, \dots, v_{n}), \\
v_{t} - v_{2,t} &= G(x, t, u, u_{1}, \dots, u_{m}, v, v_{1}, \dots, v_{n}),
\end{aligned}
\right.
\end{equation}
with $m,n\geq2$, which describes pseudospherical or spherical surfaces. To this end, we assume the associated functions $f_{ij}$ depend on $(x,t,u,u_1,\dots,u_m,v,v_1,\dots,v_n)$ and the following generic condition
\begin{equation}\label{3.2}
(F_{u_m}^2 + G_{u_m}^2) (F_{v_n}^2 + G_{v_n}^2) \neq 0,
\end{equation}
up to a subset of measure zero. This condition is not particularly restrictive and ensures that the system \eqref{3.1} depends on $u_m$ and $v_n$. In other words, up to a change of variables, the system \eqref{3.1} cannot be reduced to one of lower order.

Motivated by the concept of a differential system describing $\eta$-pseudospherical or $\eta$-spherical surfaces, as introduced by  Ding and Tenenblat \cite{ding2002}, we further assume that at least one of the functions $f_{i1}$, for $i=1,2,3$, satisfies $f_{i1}=\eta\in\mathbb{R}$. Observe that the structure equations \eqref{2.10} remain invariant under the transformation
\begin{equation}\label{3.3}
 \tilde{\omega}_{1} \rightarrow \omega_{2},\quad \tilde{\omega}_{2} \rightarrow \omega_{1},\quad \tilde{\omega}_{3} \rightarrow -\omega_{3}.
\end{equation}
Consequently, the case $f_{11}=\eta$ can be transformed into the case $f_{21}=\eta$. Although both cases yield the same systems describing pseudospherical or spherical surfaces, their associated linear problems may be different.

\subsection{Classification Theorems}

\indent \indent In this subsection, the main classification results are summarized successively in Theorem 3.4-3.7. These theorems rely on Lemma 3.1 below, which establishes existence conditions for the associated functions $f_{ij}$ and $F$, $G$, guaranteing the corresponding equation that can describe pseudospherical or spherical surfaces.

\begin{Lemma}
The necessary and sufficient conditions for a system of partial differential equations \eqref{3.1} to describe a pseudospherical surface ($\delta = 1$) or spherical surface ($\delta = -1$), with associated functions $f_{ij}=f_{ij}(x,t,u,u_1,\dots,u_m,v,v_1,\dots,v_n)$, are given by
\begin{equation}\label{3.4}
f_{i1,u}+f_{i1,u_2}=0,\quad f_{i1,v}+f_{i1,v_2}=0,\quad 1\leq i\leq 3,
\end{equation}
\begin{equation}\label{3.5}
f_{i1,u_k}=f_{i1,v_l}=0,\quad f_{i2,u_m}=f_{i2,v_n}=0,\quad k=1,3,4\dots,m,\quad l=1,3,4,\dots,n,
\end{equation}
\begin{equation}\label{3.6}
\begin{vmatrix}
f_{11,u} & f_{11,v} \\
f_{21,u} & f_{21,v}
\end{vmatrix}^2
+
\begin{vmatrix}
f_{21,u} & f_{21,v} \\
f_{31,u} & f_{31,v}
\end{vmatrix}^2
+
\begin{vmatrix}
f_{11,u} & f_{11,v} \\
f_{31,u} & f_{31,v}
\end{vmatrix}^2
\neq 0,
\end{equation}
\begin{equation}\label{3.7}
-f_{11,t}-f_{11,u}F-f_{11,v}G+D_x f_{12}-f_{31}f_{22}+f_{32}f_{21}=0,
\end{equation}
\begin{equation}\label{3.8}
-f_{12,t}-f_{21,u}F-f_{21,v}G+D_x f_{22}-f_{11}f_{32}+f_{12}f_{31}=0,
\end{equation}
\begin{equation}\label{3.9}
-f_{31,t}-f_{31,u}F-f_{31,v}G+D_x f_{32}-\delta f_{11}f_{22}+\delta f_{12}f_{21}=0,
\end{equation}
\begin{equation}\label{3.10}
f_{11}f_{22}-f_{12}f_{21}\neq 0.
\end{equation}
\end{Lemma}

\begin{Proof}
Let $u(x,t)$ and $v(x,t)$ be smooth solutions to system \eqref{3.1}. Then, from \eqref{2.19}, we obtain
\begin{equation}\label{3.11}
\begin{aligned}
& du\wedge dx=-F dx \wedge dt+du_2 \wedge dx, \quad & du_k & \wedge dt=u_{k+1} dx \wedge dt,\quad & 0\leq k\leq m-1,\\
& dv\wedge dx=-G dx \wedge dt+dv_2 \wedge dx,\quad & dv_l &\wedge dt=v_{l+1} dx \wedge dt,\quad & 0\leq l\leq n-1.
\end{aligned}
\end{equation}
Since the associated functions $f_{ij}$ depend on $(x,t,u,u_1,\dots,u_m,v,v_1,\dots,v_n)$, the exterior derivatives of the 1-forms $\omega_i$ are
\begin{equation}\label{3.12}
 \begin{split}
d \omega_{i}=&\left(f_{i2,x}-f_{i1,t}-f_{i1,u}F-f_{i1,v}G+\sum_{k=0}^{m-1}u_{k+1}f_{i2,u_k}+\sum_{l=0}^{n-1}v_{l+1}f_{i2,v_l}\right)dx \wedge dt\\
&+(f_{i1,u}+f_{i1,u_2})\,du_{2}\wedge dx+\sum_{\substack{k=1\\k\neq 2}}^{m} f_{i1,u_k}\,du_k\wedge dx+(f_{i1,v}+f_{i1,v_2})\,dv_{2}\wedge dx\\
&+\sum_{\substack{l=1\\l\neq 2}}^{n} f_{i1,v_l}\,dv_l\wedge dx+f_{i2,u_m}\,du_{m}\wedge dt+f_{i2,v_n}\,dv_{n}\wedge dt.
\end{split}
\end{equation}
Requiring that the 1-forms $\omega_1$, $\omega_2$, $\omega_3$ satisfy the structure equations \eqref{2.10}, and setting the coefficients of all independent 2-forms to zero, we deduce conditions \eqref{3.4}-\eqref{3.5} along with
\begin{equation}\label{3.13}
f_{12,x}-f_{11,t}-f_{11,u}F-f_{11,v}G+\sum_{k=0}^{m-1}u_{k+1}f_{12,u_k}+\sum_{l=0}^{n-1}v_{l+1}f_{12,v_l}-f_{31}f_{22}+f_{32}f_{21}=0,
\end{equation}
\begin{equation}\label{3.14}
f_{22,x}-f_{12,t}-f_{21,u}F-f_{21,v}G+\sum_{k=0}^{m-1}u_{k+1}f_{22,u_k}+\sum_{l=0}^{n-1}v_{l+1}f_{22,v_l}-f_{11}f_{32}+f_{12}f_{31}=0,
\end{equation}
\begin{equation}\label{3.15}
f_{32,x}-f_{31,t}-f_{31,u}F-f_{31,v}G+\sum_{k=0}^{m-1}u_{k+1}f_{32,u_k}+\sum_{l=0}^{n-1}v_{l+1}f_{32,v_l}-\delta f_{11}f_{22}+\delta f_{12}f_{21}=0.
\end{equation}
Expressing these in terms of total derivatives with respect to $x$, we get equations \eqref{3.7}-\eqref{3.9}. The constraint \eqref{3.6} is necessary to determine $F$ and $G$ uniquely from equations \eqref{3.13}-\eqref{3.15}.  Moreover, condition \eqref{3.10} ensures the existence of a metric defined on an open subset of $\mathbb{R}^2$.

The converse follows by direct computation.
$\hfill\square$
\end{Proof}

\begin{Corollary}
Under the conditions of Lemma 3.1, the associated functions $f_{i1}$, for $i=1,2,3$, are differentiable in the variables $(x,t,u - u_2,v-v_2)$, and satisfy $f_{i1,u} = -f_{i2,u_2} \neq 0$ and $f_{i1,v} = -f_{i2,v_2} \neq 0$.
\end{Corollary}

\begin{Corollary}
For the system \eqref{3.1} describing pseudospherical surfaces ($\delta = 1$) or spherical surfaces ($\delta = -1$) with associated functions $f_{ij}$ satisfying \eqref{3.4}-\eqref{3.10}, it is necessary that
\begin{equation}\label{3.16}
\left\{
\begin{aligned}
F &= F_1+F_2 u_m+F_3 v_n, \\
G &= G_1+G_2 u_m+G_3 v_n,
\end{aligned}
\right.
\end{equation}
where $F_p$, $G_p$, for $p=1,2,3$, are smooth functions of the variables $(x,t,u,u_1,\dots,u_m,v,v_1,\dots,v_n)$.
\end{Corollary}

\begin{Proof}
Differentiating equations \eqref{3.7}-\eqref{3.9} twice with respect to $u_m$ and $v_n$, and making use of condition \eqref{3.6}, we conclude that $F$ and $G$ must be linear in $u_m$ and $v_n$.
$\hfill\square$
\end{Proof}

The following theorems provide a classification of the system \eqref{3.1} in terms of four arbitrary smooth functions satisfying certain generic conditions. The results are presented separately for pseudospherical or spherical surfaces, with complete proofs provided for each case.

\begin{Theorem}
A system of partial differential equations of the form \eqref{3.1}, satisfying \eqref{3.2}, describes pseudospherical surfaces ($\delta = 1$) or spherical surfaces ($\delta = -1$) with associated functions $f_{ij}$, satisfying \eqref{3.4}-\eqref{3.10}, and with $f_{21}=\eta\in\mathbb{R}$, if and only if it can be written as
\begin{equation}\label{3.17}
\begin{pmatrix}
u_{t}\\
v_{t}
\end{pmatrix}
-
\begin{pmatrix}
u_{2,t}\\
v_{2,t}
\end{pmatrix}
=\frac{1}{W}
\begin{pmatrix}
h_v & -g_v \\
-h_u & g_u
\end{pmatrix}
\begin{pmatrix}
-g_t+D_x L-hM+\eta N \\
-h_t+D_x N-\delta gM+\delta\eta L
\end{pmatrix},
\end{equation}
where $g=g(x,t,u,u_2,v,v_2)$, $h=h(x,t,u,u_2,v,v_2)$ are smooth functions such that $W:=g_u h_v - g_v h_u\neq 0$,  $L=L(x,t,u,u_1,\dots,u_{m-1},v,v_1,\dots,v_{n-1})$, $M = M(x, t, u, u_1, \dots, u_{m-2},v,v_1, \dots, v_{n-2})$ are smooth functions satisfying $gM-\eta L\neq 0$ and the generic condition
\begin{equation}\label{3.18}
(L_{u_{m-1}}^2+N_{u_{m-1}}^2)(L_{v_{n-1}}^2+N_{v_{n-1}}^2)\neq 0,
\end{equation}
with
\begin{equation}\label{3.19}
N:=\frac{1}{g}(D_x M+hL).
\end{equation}
Moreover, the associated functions $f_{ij}$ are given by
\begin{equation}\label{3.20}
\begin{aligned}
f_{11} &=g,\quad & f_{12} &=L,\\
f_{21} &=\eta,\quad & f_{22} &=M,\\
f_{31} &=h,\quad & f_{32} &=N.
\end{aligned}
\end{equation}
\end{Theorem}

\begin{Proof}
By Corollary 3.2, the functions $f_{11}$ and $f_{31}$ depend on the variables $(x,t,u - u_2,v-v_2)$. According to Lemma 3.1, the associated functions $f_{ij}$ satisfy \eqref{3.4}-\eqref{3.10}, and each $f_{i2}$ depends on $(x,t,u,u_1,\dots, u_m,v,v_1,\dots, v_n)$ for $i=1,2,3$. Condition \eqref{3.6} simplifies to
\begin{equation}\label{3.21}
W=f_{11,u}f_{31,v}-f_{11,v}f_{31,u}\neq 0.
\end{equation}
Under this condition, equations \eqref{3.7} and \eqref{3.9} can be expressed in matrix form as
\begin{equation}\label{3.22}
\begin{pmatrix}
F\\
G
\end{pmatrix}
=\frac{1}{W}
\begin{pmatrix}
f_{31,v} & -f_{11,v} \\
-f_{31,u} & f_{11,u}
\end{pmatrix}
\begin{pmatrix}
-f_{11,t}+D_x f_{12}-f_{31}f_{22}+\eta f_{32} \\
-f_{31,t}+D_x f_{32}-\delta f_{11}f_{22}+\delta\eta f_{12}
\end{pmatrix}.
\end{equation}
Furthermore, equation \eqref{3.8} becomes
\begin{equation}\label{3.23}
D_x f_{22}-f_{11}f_{32}+f_{12}f_{31}=0.
\end{equation}
Differentiating equation \eqref{3.23} with respect to $u_m$ and $v_n$ yields
\begin{equation}\label{3.24}
f_{22,u_{m-1}}=f_{22,v_{n-1}}=0,
\end{equation}
which implies that $f_{22}$ depends only on $(x,t,u,u_1,\dots, u_{m-1},v,v_1,\dots, v_{n-1})$. Since $f_{11}\neq0$, it follows from equation \eqref{3.23} that
\begin{equation}\label{3.25}
f_{32}=\frac{1}{f_{11}}(D_x f_{22}+f_{12}f_{31}).
\end{equation}
Let $g=f_{11}$, $h=f_{31}$, $L=f_{12}$, $M=f_{22}$, $N=f_{32}$. Then condition \eqref{3.10} is equivalent to $gM-\eta L\neq 0$, and the generic condition \eqref{3.2} reduces to \eqref{3.18}.

The converse follows by direct computation.
$\hfill\square$
\end{Proof}

\begin{Theorem}
A system of partial differential equations of the form \eqref{3.1}, satisfying \eqref{3.2}, describes pseudospherical surfaces ($\delta = 1$) or spherical surfaces ($\delta = -1$) with associated functions $f_{ij}$, satisfying \eqref{3.4}-\eqref{3.10}, and with $f_{31}=\eta\in\mathbb{R}$ if and only if it can be written as
\begin{equation}\label{3.26}
\begin{pmatrix}
u_{t}\\
v_{t}
\end{pmatrix}
-
\begin{pmatrix}
u_{2,t}\\
v_{2,t}
\end{pmatrix}
=\frac{1}{W}
\begin{pmatrix}
h_v & -g_v \\
-h_u & g_u
\end{pmatrix}
\begin{pmatrix}
-g_t+D_x L-\eta N+hM \\
-h_t+D_x N-gM+\eta L
\end{pmatrix},
\end{equation}
where $g=g(x,t,u,u_2,v,v_2)$, $h=h(x,t,u,u_2,v,v_2)$ are smooth functions such that $W:=g_u h_v - g_v h_u\neq 0$, $L=L(x,t,u,u_1,\dots,u_{m-1},v,v_1,\dots,v_{n-1})$ is a smooth function, and $M=M(x,t,u,u_1,\dots,u_{m-2},v,v_1,\dots,v_{n-2})$ is a non-constant smooth function satisfying the generic condition
\begin{equation}\label{3.27}
(L_{u_{m-1}}^2+N_{u_{m-1}}^2)(L_{v_{n-1}}^2+N_{v_{n-1}}^2)\neq 0,
\end{equation}
with
\begin{equation}\label{3.28}
N:=\frac{1}{g}(\delta D_x M+hL).
\end{equation}
Moreover, the associated functions $f_{ij}$ are given by
\begin{equation}\label{3.29}
\begin{aligned}
f_{11} &=g,\quad & f_{12} &=L,\\
f_{21} &=h,\quad & f_{22} &=N,\\
f_{31} &=\eta,\quad & f_{32} &=M.
\end{aligned}
\end{equation}
\end{Theorem}

\begin{Proof}
It follows from Corollary 3.2 that the functions $f_{11}$ and $f_{21}$ depend on the variables $(x,t,u - u_2,v-v_2)$. By Lemma 3.1, the associated functions $f_{ij}$ satisfy \eqref{3.4}-\eqref{3.10}, and each $f_{i2}$ depends on $(x,t,u,u_1,\dots, u_m,v,v_1,\dots, v_n)$ for $i=1,2,3$. Condition \eqref{3.6} becomes
\begin{equation}\label{3.30}
W=f_{11,u}f_{21,v}-f_{11,v}f_{21,u}\neq 0.
\end{equation}
Under this condition, equations \eqref{3.7} and \eqref{3.8} are equivalent to
\begin{equation}\label{3.31}
\begin{pmatrix}
F\\
G
\end{pmatrix}
=\frac{1}{W}
\begin{pmatrix}
f_{21,v} & -f_{11,v} \\
-f_{21,u} & f_{11,u}
\end{pmatrix}
\begin{pmatrix}
-f_{11,t}+D_x f_{12}-\eta f_{22}+f_{21} f_{32} \\
-f_{21,t}+D_x f_{22}- f_{11}f_{32}+\eta f_{12}
\end{pmatrix}.
\end{equation}
Furthermore, equation \eqref{3.9} simplifies to
\begin{equation}\label{3.32}
D_x f_{32}-\delta f_{11}f_{22}+\delta f_{12}f_{21}=0.
\end{equation}
Differentiating equation \eqref{3.32} with respect to $u_m$ and $v_n$ gives
\begin{equation}\label{3.33}
f_{32,u_{m-1}}=f_{32,v_{n-1}}=0.
\end{equation}
Thus, the function $f_{32}$ does not depend on $u_{m-1}$ and $v_{n-1}$. In addition, since $f_{11}\neq0$, it follows from \eqref{3.32} that
\begin{equation}\label{3.34}
f_{22}=\frac{1}{f_{11}}(\delta D_x f_{32}+f_{12}f_{21}).
\end{equation}
Let $g=f_{11}$, $h=f_{21}$, $L=f_{12}$, $N=f_{22}$, $M=f_{32}$. From condition \eqref{3.10}, we derive $D_x M= 0$, implying that $M$ cannot be a constant.  Finally, the generic condition \eqref{3.2} reduces to \eqref{3.27}.

The converse follows by direct computation.
$\hfill\square$
\end{Proof}

Motivated by the Song-Qu-Qiao system \cite{song2011}, the following two theorems address third-order systems of partial differential equations of the form
\begin{equation}\label{3.35}
\left\{
\begin{aligned}
u_{t} - u_{2,t} &= A_1(u,u_1,v,v_1)u_3+B_1(u,u_1,u_2,v,v_1,v_2), \\
v_{t} - v_{2,t} &= A_2(u,u_1,v,v_1)u_3+B_2(u,u_1,u_2,v,v_1,v_2),
\end{aligned}
\right.
\end{equation}
where $A_p(u,u_1,v,v_1)\neq 0$, $B_p(u,u_1,u_2,v,v_1,v_2)$, $p=1,2$, are smooth functions. These results will determine whether such systems describe pseudospherical or spherical surfaces, under the assumption that the associated functionss $f_{ij}$ do not depend explicitly on the independent variables $x$ and $t$. As in previous classifications, two cases are considered: $f_{21}=\eta$ and $f_{31}=\eta$. Note that in both cases, the third-order coefficients must be same, namely, $A_1=A_2$.

\begin{Theorem}
A third-order system of partial differential equations of the form \eqref{3.35} describes pseudospherical surfaces ($\delta = 1$) or spherical surfaces ($\delta = -1$) with associated functions $f_{ij}$ satisfying \eqref{3.4}-\eqref{3.10}, and with $f_{21}=\eta\in\mathbb{R}$, if and only if $A_1=A_2:=A(u,u_1,v,v_1)$, and the system takes the form
\begin{equation}\label{3.36}
\begin{aligned}
\begin{pmatrix}
u_{t}\\
v_{t}
\end{pmatrix}
-
\begin{pmatrix}
u_{2,t}\\
v_{2,t}
\end{pmatrix}
=&A
\begin{pmatrix}
u_3 \\
v_3
\end{pmatrix}
-\frac{D_x A}{W}
\begin{pmatrix}
gh_v-hg_v \\
-gh_u+hg_u
\end{pmatrix}
-A
\begin{pmatrix}
u_1 \\
v_1
\end{pmatrix}
+\frac{1}{W}
\begin{pmatrix}
h_v D_x L_1-g_v D_x N_1 \\
-h_u D_x L_1+g_u D_x N_1
\end{pmatrix}\\
&-\frac{\eta A+M}{2W}
\begin{pmatrix}
(h^2-\delta g^2)_v\\
(\delta g^2-h^2)_u
\end{pmatrix}
+\frac{\eta}{W}
\begin{pmatrix}
(\delta gL_1-hN_1)_{v_2} \\
(hN_1-\delta gL_1)_{u_2}
\end{pmatrix},
\end{aligned}
\end{equation}
where $g$, $h$ are smooth functions of $(u-u_2, v-v_2)$ such that $W:= g_u h_v- g_v h_u\neq 0$, $L_1$, $ N_1$, $M$ are smooth functions of $(u,u_1,v,v_1)$ satisfying $L_1\neq \frac{g}{\eta}(M+\eta A)$ and
\begin{equation}\label{3.37}
(gN_1-hL_1)_{u_2 v_1}=(gN_1-hL_1)_{u_1 v_2}.
\end{equation}
Moreover, system \eqref{3.36} is the integrability condition of the linear problem
\begin{equation}\label{3.38}
\phi_x=X\phi,\quad \phi_t=T\phi,
\end{equation}
with $\phi=(u,v)^{T}$, where
\begin{equation}\label{3.39}
X=\frac{1}{2}
\begin{pmatrix}
\eta & g-h\\
g+h & -\eta
\end{pmatrix}, \quad
T=\frac{1}{2}
\begin{pmatrix}
M & -A(g-h)+L_1-N_1\\
-A(g+h)+L_1+N_1 & -M
\end{pmatrix},
\end{equation}
if $\delta=1$, and
\begin{equation}\label{3.40}
X=\frac{1}{2}
\begin{pmatrix}
i\eta & g+ih\\
-g+ih & -i\eta
\end{pmatrix}, \quad
T=\frac{1}{2}
\begin{pmatrix}
iM & -A(g+ih)+L_1+N_1\\
A(g-ih)-L_1+iN_1 & -iM
\end{pmatrix},
\end{equation}
if $\delta=-1$.
\end{Theorem}

\begin{Proof}
Assume the system of partial differential equations of the form \eqref{3.35} describes pseudospherical surfaces ($\delta = 1$) or spherical surfaces ($\delta = -1$) with $f_{21}=\eta$. According to Theorem 3.4 with $m=n=3$, the system is expressed as
\begin{equation}\label{3.41}
\begin{pmatrix}
u_{t}\\
v_{t}
\end{pmatrix}
-
\begin{pmatrix}
u_{2,t}\\
v_{2,t}
\end{pmatrix}
=\frac{1}{W}
\begin{pmatrix}
h_v & -g_v \\
-h_u & g_u
\end{pmatrix}
\begin{pmatrix}
\displaystyle\sum_{k=0}^{2} (L_{u_k} u_{k+1}+L_{v_k} v_{k+1})-hM+\eta N \\
\displaystyle\sum_{k=0}^{2} (N_{u_k} u_{k+1}+N_{v_k} v_{k+1})-\delta gM+\delta\eta L
\end{pmatrix},
\end{equation}
where the functions $g=g(u-u_2,v-v_2)$, $h=h(u-u_2,v-v_2)$, $L=L(u,u_1,u_2,v,v_1,v_2)$, $N=N(u,u_1,u_2,v,v_1,v_2)$ and $M=M(u,u_1,v,v_1)$ are smooth and satisfy the following conditions: $W:=g_u h_v - g_v h_u\neq 0$, $gM-\eta L\neq 0$ and $(L_{u_2}^2+N_{u_2}^2)(L_{v_2}^2+N_{v_2}^2)\neq 0$. Furthermore, the functions $L$ and $M$ are constrained by
\begin{equation}\label{3.42}
\sum_{k=0}^{1} (M_{u_k} u_{k+1}+M_{v_k} v_{k+1})+hL-Ng=0.
\end{equation}

Comparing the coefficient of $u_3$ and $v_3$ in \eqref{3.35} and \eqref{3.41}, we obtain
\begin{equation}\label{3.43}
\frac{1}{W}
\begin{pmatrix}
h_v & -g_v \\
-h_u & g_u
\end{pmatrix}
\begin{pmatrix}
L_{u_2} & L_{v_2}  \\
N_{u_2} & N_{v_2}
\end{pmatrix}
=
\begin{pmatrix}
A_1 & 0 \\
0 & A_2
\end{pmatrix},
\end{equation}
which implies
\begin{equation}\label{3.44}
\begin{pmatrix}
L_{u_2} & L_{v_2}  \\
N_{u_2} & N_{v_2}
\end{pmatrix}
=
\begin{pmatrix}
g_u & g_v \\
h_u & h_v
\end{pmatrix}
\begin{pmatrix}
A_1 & 0 \\
0 & A_2
\end{pmatrix}.
\end{equation}
Taking the mixed derivatives of $L$ and $N$ with respect to $u_2$ and $v_2$, we find
\begin{equation}\label{3.45}
g_{u_2 v_2}(A_1-A_2)=0,\quad h_{u_2 v_2}(A_1-A_2)=0.
\end{equation}
Now, there are two cases to consider, either $A_1=A_2$ or $A_1\neq A_2$. we claim that $A_1=A_2$ must hold. Suppose, for contradiction, that $A_1\neq A_2$, then $g_{u_2 v_2}=h_{u_2 v_2}=0$, and hence
\begin{equation}\label{3.46}
g=g_1(u-u_2)+g_2(v-v_2),\quad h=h_1(u-u_2)+h_2(v-v_2).
\end{equation}
From equation \eqref{3.44}, we deduce
\begin{equation}\label{3.47}
\begin{aligned}
L&=-(g_1(u-u_2) A_1+g_2(v-v_2) A_2+L_1(u,u_1,v,v_1)), \\
N&=-(h_1(u-u_2) A_1+h_2(v-v_2) A_2+N_1(u,u_1,v,v_1)).
\end{aligned}
\end{equation}
By taking the mixed derivatives of equation \eqref{3.42} with respect to $u_2$ and $v_2$, we conclude
\begin{equation}\label{3.48}
(A_1-A_2)(h_1^{\prime}g_2^{\prime}-h_2^{\prime}g_1^{\prime})=0.
\end{equation}
Since $W:= g_u h_v- g_v h_u\neq 0$, it follows that $h_1^{\prime}g_2^{\prime}-h_2^{\prime}g_1^{\prime}\neq0$, which leads to a contradiction. Thus, the case $A_1\neq A_2$ does not occur.

Therefore, we must have $A_1=A_2=A(u,u_1,v,v_1)$. Moreover, from equations \eqref{3.42} and \eqref{3.44}, we derive
\begin{equation}\label{3.49}
L=-Ag+L_1(u,u_1,v,v_1),\quad N=-Ag+N_1(u,u_1,v,v_1),
\end{equation}
\begin{equation}\label{3.50}
M_u u_1+M_{u_1} u_2+M_v v_1+M_{v_1} v_2+hL_1-Ng_1=0.
\end{equation}
Differentiating equation \eqref{3.50} with respect to $u_2$ and $v_2$, and applying the compatibility condition $M_{u_1 v_1}=M_{v_1 u_1}$, we get equation \eqref{3.37}. The condition $gM-\eta L\neq0$ gives $L_1\neq \frac{g}{\eta}(M+\eta A)$. Finally, the system \eqref{3.41} reduces to \eqref{3.36}.
$\hfill\square$
\end{Proof}

\begin{Theorem}
A third-order system of partial differential equations of the form \eqref{3.35} describes pseudospherical surfaces ($\delta = 1$) or spherical surfaces ($\delta = -1$) with associated functions $f_{ij}$ satisfying \eqref{3.4}-\eqref{3.10}, and with $f_{31}=\eta\in\mathbb{R}$, if and only if $A_1=A_2:=A(u,u_1,v,v_1)$, and the system takes the form
\begin{equation}\label{3.51}
\begin{aligned}
\begin{pmatrix}
u_{t}\\
v_{t}
\end{pmatrix}
-
\begin{pmatrix}
u_{2,t}\\
v_{2,t}
\end{pmatrix}
=&A
\begin{pmatrix}
u_3 \\
v_3
\end{pmatrix}
-\frac{D_x A}{W}
\begin{pmatrix}
gh_v-hg_v \\
-gh_u+hg_u
\end{pmatrix}
-A
\begin{pmatrix}
u_1 \\
v_1
\end{pmatrix}
+\frac{1}{W}
\begin{pmatrix}
h_v D_x L_1-g_v D_x N_1 \\
-h_u D_x L_1+g_u D_x N_1
\end{pmatrix}\\
&+\frac{\eta A+M}{2W}
\begin{pmatrix}
(h^2+g^2)_v\\
-(h^2+g^2)_u
\end{pmatrix}
-\frac{\eta}{W}
\begin{pmatrix}
-( hN_1+gL_1)_{v_2} \\
(hN_1+gL_1)_{u_2}
\end{pmatrix},
\end{aligned}
\end{equation}
where $g$, $h$ are smooth functions of $(u-u_2, v-v_2)$ such that $W:= g_u h_v- g_v h_u\neq 0$, and $L_1$, $ N_1$, $M$ are smooth functions of $(u,u_1,v,v_1)$ with $M$ non-constant, satisfying
\begin{equation}\label{3.52}
\delta D_x M+hL_1-gN_1=0.
\end{equation}
Moreover, system \eqref{3.51} is the integrability condition of the linear problem
\begin{equation}\label{3.53}
\phi_x=X\phi,\quad \phi_t=T\phi,
\end{equation}
with $\phi=(u,v)^{T}$, where
\begin{equation}\label{3.54}
X=\frac{1}{2}
\begin{pmatrix}
h & g-\eta\\
g+\eta & -h
\end{pmatrix}, \quad
T=\frac{1}{2}
\begin{pmatrix}
-Ah+N_1 & -Ag+L_1-M\\
-Ag+L_1+M & Ah-N_1
\end{pmatrix},
\end{equation}
if $\delta=1$, and
\begin{equation}\label{3.55}
X=\frac{1}{2}
\begin{pmatrix}
ih & g+i\eta\\
-g+i\eta & -ih
\end{pmatrix}, \quad
T=\frac{1}{2}
\begin{pmatrix}
-iAh+iN_1 & -Ag+L_1+iM\\
Ag-L_1+iM & iAh-iN_1
\end{pmatrix},
\end{equation}
if $\delta=-1$.
\end{Theorem}

The proof follows the same steps as in the proof of Theorem 3.6 and is therefore omitted.

\subsection{Examples}

\indent \indent In this subsection, we provide several examples of systems of partial differential equations describing pseudospherical or spherical surfaces of type \eqref{3.1}. We include well-known examples such as Song-Qu-Qiao system \cite{song2011,tian2013,chang2016,qiao2015a}, the two-component CH system with cubic nonlinearity \cite{qiao2013,qiao2015b,xia2015a} and the modified CH-type system.

\begin{Example}
The Song-Qu-Qiao system introduced in \cite{song2011}
\begin{equation}\label{3.56}
\left\{
\begin{aligned}
u_{t} - u_{2t} &= [(u-u_{2})(u_1 v_1 -uv+uv_1-u_1 v)]_x, \\
v_{t} - v_{2t} &= [(v-v_{2})(u_1 v_1 -uv+uv_1-u_1 v)]_x,
\end{aligned}
\right.
\end{equation}
describes \textbf{pseudospherical surfaces}, with associated functions
\begin{equation}\label{3.57}
\begin{aligned}
f_{11} &=\eta\left[(u-u_{2})e^{(\eta-1)x}+(v-v_{2})e^{-(\eta-1)x}\right],\\
f_{12} &=\eta Q\left[(u-u_{2})e^{(\eta-1)x}+(v-v_{2})e^{-(\eta-1)x}\right]+\frac{1}{2\eta}\left[(u+u_{1})e^{(\eta-1)x}+(v-v_{1})e^{-(\eta-1)x}\right],\\
f_{21} &=\eta,\\
f_{22} &=\frac{1}{2\eta^2}+Q,\\
f_{31} &=-\eta\left[(u-u_{2})e^{(\eta-1)x}-(v-v_{2})e^{-(\eta-1)x}\right],\\
f_{32} &=-\eta Q\left[(u-u_{2})e^{(\eta-1)x}-(v-v_{2})e^{-(\eta-1)x}\right]-\frac{1}{2\eta}\left[(u+u_{1})e^{(\eta-1)x}-(v-v_{1})e^{-(\eta-1)x}\right].
\end{aligned}
\end{equation}
Moreover, system \eqref{3.56} is the integrability condition of the linear problem
\begin{equation}\label{3.58}
\begin{aligned}
\phi_x &=\frac{1}{2}
\begin{pmatrix}
\eta & 2\eta (u-u_{2})e^{(\eta-1)x}\\[0.4em]
2\eta (v-v_{2})e^{-(\eta-1)x} & -\eta
\end{pmatrix} \phi,\\
\phi_t&=\frac{1}{2}
\begin{pmatrix}
\frac{1}{2\eta^2}+Q & \left[2\eta Q (u-u_{2})+\frac{1}{\eta}(u+u_1)\right]e^{(\eta-1)x}\\[0.4em]
\left[2\eta Q (v-v_{2})+\frac{1}{\eta}(v-v_1)\right]e^{-(\eta-1)x} &  -\frac{1}{2\eta^2}-Q
\end{pmatrix}\phi,
\end{aligned}
\end{equation}
where $\eta\neq 0$ is a real parameter, $\phi=(\phi_1,\phi_2)^{T}$ and $Q=u_1 v_1 -uv+uv_1-u_1 v$.
\end{Example}

\begin{Remark}
The system \eqref{3.56} is reduced to the modified CH equation \cite{fokas1995,olver1996,qiao2006,fuchssteiner1996} as $v=-u$,
\begin{equation}\label{3.59}
m_t=[m(u^2-u_1^2)]_x, \quad m=u-u_{2}.
\end{equation}
\end{Remark}

\begin{Example}
The two-component CH system with cubic nonlinearity proposed by Xia and Qiao in \cite{qiao2013}
\begin{equation}\label{3.60}
\left\{
\begin{aligned}
u_{t} - u_{2t} &= \frac{1}{2}[(u-u_{2})(uv-u_1 v_1)]_x-\frac{1}{2}(u-u_{2})(u v_1 -u_1 v), \\
v_{t} - v_{2t} &= \frac{1}{2}[(v-v_{2})(uv-u_1 v_1)]_x+\frac{1}{2}(v-v_{2})(u v_1 -u_1 v),
\end{aligned}
\right.
\end{equation}
describes \textbf{pseudospherical surfaces}, with associated functions
\begin{equation}\label{3.61}
\begin{aligned}
f_{11} &=\frac{1}{2}\eta[(u-u_{2})-(v-v_{2})],\\
f_{12} &=\frac{1}{4}\eta(uv-u_1 v_1)[(u-u_{2})-(v-v_{2})]+\frac{1}{2\eta}[(u-u_{1})-(v+v_{1})],\\
f_{21} &=-1,\\
f_{22} &=-\frac{1}{\eta^2}-\frac{1}{2}(uv-u_1 v_1+uv_1-u_1 v),\\
f_{31} &=-\frac{1}{2}\eta[(u-u_{2})+(v-v_{2})],\\
f_{32} &=-\frac{1}{4}\eta(uv-u_1 v_1)[(u-u_{2})+(v-v_{2})]-\frac{1}{2\eta}[(u-u_{1})+(v+v_{1})].
\end{aligned}
\end{equation}
Moreover, system \eqref{3.60} is the integrability condition of the linear problem
\begin{equation}\label{3.62}
\begin{aligned}
\phi_x &=\frac{1}{2}
\begin{pmatrix}
-1 & \eta (u-u_{2})\\[0.4em]
-\eta (v-v_{2}) & 1
\end{pmatrix} \phi,\\
\phi_t&=\frac{1}{2}
\begin{pmatrix}
-\frac{1}{\eta^2}-\frac{1}{2}(uv-u_1 v_1+uv_1-u_1 v) & \frac{1}{2}\eta (uv-u_1 v_1) (u-u_{2})+\frac{1}{\eta}(u-u_1)\\[0.4em]
-\frac{1}{2}\eta (uv-u_1 v_1) (v-v_{2})-\frac{1}{\eta}(v+v_1) &  \frac{1}{\eta^2}+\frac{1}{2}(uv-u_1 v_1+uv_1-u_1 v)
\end{pmatrix}\phi,
\end{aligned}
\end{equation}
where $\eta\neq 0$ is a real parameter and $\phi=(\phi_1,\phi_2)^{T}$.
\end{Example}

\begin{Remark}
The system \eqref{3.60} reduces to the modified CH equation \eqref{3.59} as $v=2u$.
\end{Remark}

\begin{Example}
The following system \cite{xia2015b}
\begin{equation}\label{3.63}
\left\{
\begin{aligned}
u_{t} - u_{2t} &= -\frac{1}{2}(u-u_{2})(u-u_1 )(v+v_1), \\
v_{t} - v_{2t} &=\frac{1}{2}(v-v_{2})(u-u_1)(v+v_1),
\end{aligned}
\right.
\end{equation}
describes \textbf{pseudospherical surfaces}, with associated functions
\begin{equation}\label{3.64}
\begin{aligned}
f_{11} &=\frac{1}{2}\eta[(v-v_{2})-(u-u_{2})], \quad & f_{12} &= \frac{1}{2\eta}[(v+v_{1})-(u-u_{1})],\\
f_{21} &=1, \quad & f_{22} &=\frac{1}{\eta^2}+\frac{1}{2}(u-u_1 )(v+v_1),\\
f_{31} &=-\frac{1}{2}\eta[(u-u_{2})+(v-v_{2})],\quad & f_{32} &=-\frac{1}{2\eta}[(u-u_{1})+(v+v_{1})].
\end{aligned}
\end{equation}
Moreover, system \eqref{3.63} is the integrability condition of the linear problem
\begin{equation}\label{3.65}
\begin{aligned}
\phi_x &=\frac{1}{2}
\begin{pmatrix}
1 & \eta (v-v_{2})\\[0.4em]
-\eta (u-u_{2}) & -1
\end{pmatrix} \phi,\\
\phi_t&=\frac{1}{2}
\begin{pmatrix}
\frac{1}{\eta^2}+\frac{1}{2}(u-u_1 )(v+v_1) & \frac{1}{\eta}(v+v_1)\\[0.4em]
-\frac{1}{\eta}(u-u_1) &  -\frac{1}{\eta^2}-\frac{1}{2}(u-u_1 )(v+v_1)
\end{pmatrix}\phi,
\end{aligned}
\end{equation}
where $\eta\neq 0$ is a real parameter and $\phi=(\phi_1,\phi_2)^{T}$.
\end{Example}

\begin{Example}
The modified CH-type system
\begin{equation}\label{3.66}
\left\{
\begin{aligned}
u_{t} - u_{2t} &=-\left[\left(\frac{1}{2}(u^2 +v^2-u_1^2-v_1^2)+(uv_1 -u_1 v)\right)(u-u_{2})\right]_x-2u_1 , \\
v_{t} - v_{2t} &= -\left[\left(\frac{1}{2}(u^2 +v^2-u_1^2-v_1^2)+(uv_1 -u_1 v)\right)(v-v_{2})\right]_x-2v_1,
\end{aligned}
\right.
\end{equation}
describes \textbf{spherical surfaces}, with associated functions
\begin{equation}\label{3.67}
\begin{aligned}
f_{11} &=-(v-v_{2}),\quad & f_{12} &=-R(v-v_{2})+v+u_1,\\
f_{21} &=1, \quad & f_{22} &=R-1,\\
f_{31} &=u-u_{2},\quad & f_{32} &=R(u-u_{2})-u+v_1.
\end{aligned}
\end{equation}
Moreover, system \eqref{3.66} is the integrability condition of the linear problem
\begin{equation}\label{3.68}
\begin{aligned}
\phi_x &=\frac{1}{2}
\begin{pmatrix}
i &  -n+im\\[0.4em]
n+im & -i
\end{pmatrix} \phi,\\
\phi_t&=\frac{1}{2}
\begin{pmatrix}
i(R-1) & -R(n-im)+v+u_1+i(v_1-u) \\[0.4em]
R(n+im)-v-u_1+i(v_1-u) & -i(R-1)
\end{pmatrix}\phi,
\end{aligned}
\end{equation}
where $\phi=(\phi_1,\phi_2)^{T}$, $m=u-u_{2},n=v-v_{2}$ and $R=-\frac{1}{2}(u^2 +v^2-u_1^2-v_1^2)-uv_1 +u_1 v$.
\end{Example}

\begin{Remark}
The system \eqref{3.66} is reduced to the cubic CH equation \cite{fokas1995,fuchssteiner1996,olver1996} as $v=u$
\begin{equation}\label{3.69}
m_t=bu_1-[m(u^2-u_1^2)]_x, \quad m=u-u_{2},
\end{equation}
where $b$ is an arbitrary constant.
\end{Remark}

\begin{Example}
The following system \cite{xia2015b}
\begin{equation}\label{3.70}
\left\{
\begin{aligned}
u_{t} - u_{2t} &= \frac{1}{2}[(u-u_{2})(uv_1 -u_1 v)]_x-\frac{1}{2}(u-u_{2})(uv -u_1 v_1), \\
v_{t} - v_{2t} &= \frac{1}{2}[(v-v_{2})(uv_1 -u_1 v)]_x+\frac{1}{2}(v-v_{2})(uv -u_1 v_1),
\end{aligned}
\right.
\end{equation}
describes \textbf{pseudospherical surfaces}, with associated functions
\begin{equation}\label{3.71}
\begin{aligned}
f_{11} &=-\frac{1}{2}\eta[(u-u_{2})-(v-v_{2})],\\
f_{12} &=-\frac{1}{4}\eta(uv_1 -u_1 v)[(u-u_{2})-(v-v_{2})]-\frac{1}{2\eta}[(u-u_{1})-(v+v_{1})],\\
f_{21} &=1,\\
f_{22} &=\frac{1}{\eta^2}+\frac{1}{2}(u-u_1)(v+v_1),\\
f_{31} &=-\frac{1}{2}\eta[(u-u_{2})+(v-v_{2})],\\
f_{32} &=-\frac{1}{4}\eta(uv_1 -u_1 v)[(u-u_{2})+(v-v_{2})]-\frac{1}{2\eta}[(u-u_{1})+(v+v_{1})].
\end{aligned}
\end{equation}
Moreover, system \eqref{3.70} is the integrability condition of the linear problem
\begin{equation}\label{3.72}
\begin{aligned}
\phi_x &=\frac{1}{2}
\begin{pmatrix}
1 & \eta (v-v_{2})\\[0.4em]
-\eta (u-u_{2}) & -1
\end{pmatrix} \phi,\\
\phi_t&=\frac{1}{2}
\begin{pmatrix}
\frac{1}{\eta^2}+\frac{1}{2}(u-u_1)(v+v_1)& \frac{1}{2}\eta (uv_1 -u_1 v) (v-v_{2})+\frac{1}{\eta}(v+v_1)\\[0.4em]
-\frac{1}{2}\eta (uv_1 -u_1 v) (u-u_{2})-\frac{1}{\eta}(u-u_1) &  -\frac{1}{\eta^2}-\frac{1}{2}(u-u_1)(v+v_1)
\end{pmatrix}\phi,
\end{aligned}
\end{equation}
where $\eta\neq 0$ is a real parameter and $\phi=(\phi_1,\phi_2)^{T}$.
\end{Example}

\section{Nonlocal symmetry}
\indent \indent In this section, we calculate nonlocal symmetry from gradient of spectral parameter for the two-component CH system with cubic nonlinearity presented in Example 3.10. By introducing an appropriate pseudo-potential, a reduced nonlocal symmetry is prolonged to an enlarged system, thereby generating a finite symmetry transformation. As a result, we obtain a nontrivial solutions for the system.

The following applications utilize the original notation, not the one in \eqref{2.19}. Therefore, the two-component CH system with cubic nonlinearity is written in the form
\begin{equation}\label{4.1}
\left\{
\begin{aligned}
m_{t}  &= \frac{1}{2}[m(uv-u_x v_x)]_x-\frac{1}{2}m(u v_x -u_x v), \\
n_{t}  &= \frac{1}{2}[n(uv-u_x v_x)]_x+\frac{1}{2}n(u v_x -u_x v), \\
m &=u-u_{xx},\quad n=v-v_{xx}.
\end{aligned}
\right.
\end{equation}
This system admits the bi-Hamiltonian structure \cite{xia2015b}
\begin{equation}\label{4.2}
\begin{pmatrix}
m_t & n_t
\end{pmatrix}^T
=\mathcal{D}_1
\begin{pmatrix}
\displaystyle\frac{\delta \mathcal{H}_2}{\delta m} & \displaystyle\frac{\delta \mathcal{H}_2}{\delta n}
\end{pmatrix}^T
=\mathcal{D}_2
\begin{pmatrix}
\displaystyle\frac{\delta \mathcal{H}_1}{\delta m}& \displaystyle\frac{\delta \mathcal{H}_1}{\delta n}
\end{pmatrix}^T ,
\end{equation}
where $\mathcal{D}_1$ and $\mathcal{D}_2$ are two compatible Hamiltonian operators
\begin{equation}\label{4.3}
\mathcal{D}_1=
\begin{pmatrix}
0 & \partial_x^2-1\\[0.4em]
1-\partial_x^2 & 0
\end{pmatrix},\quad
\mathcal{D}_2=
\begin{pmatrix}
\displaystyle\partial_x m \partial_x^{-1}m\partial_x-m\partial_x^{-1} m& \displaystyle \partial_x m \partial_x^{-1}n\partial_x+m\partial_x^{-1} n\\[0.4em]
\displaystyle \partial_x n \partial_x^{-1}m\partial_x+n\partial_x^{-1} m & \displaystyle \partial_x n \partial_x^{-1}n\partial_x-n\partial_x^{-1} n
\end{pmatrix},
\end{equation}
and
\begin{equation}\label{4.4}
\mathcal{H}_1=\frac{1}{2}\int(uv+u_x v_x)dx,\quad \mathcal{H}_2=\frac{1}{4}\int(u^2 v_x+u_x^2 v_x-2uu_x v)ndx.
\end{equation}
Its linear problem \eqref{3.62} is formulated as
\begin{subequations}\label{4.5}
\begin{align}
\begin{pmatrix}
\phi_1\\
\phi_2
\end{pmatrix}_x
&=M
\begin{pmatrix}
\phi_1\\
\phi_2
\end{pmatrix},
&M&=
\begin{pmatrix}
- \frac{1}{2} &  \frac{1}{2}\eta m\\[0.4em]
- \frac{1}{2}\eta n &  \frac{1}{2}
\end{pmatrix},\label{4.5a}\\
\begin{pmatrix}
\phi_1\\
\phi_2
\end{pmatrix}_t
&=N
\begin{pmatrix}
\phi_1\\
\phi_2
\end{pmatrix},
&N&=
\begin{pmatrix}
-\alpha & \frac{1}{4}\eta m\beta+\frac{1}{2\eta}(u-u_x)\\[0.4em]
-\frac{1}{4}\eta n\beta -\frac{1}{2\eta}(v+v_x) &  \alpha
\end{pmatrix},\label{4.5b}
\end{align}
\end{subequations}
where $\alpha=\frac{1}{2\eta^2}+\frac{1}{4}(uv-u_x v_x+uv_x-u_x v)$, $\beta=uv-u_x v_x$ and $\eta\neq0$ is the spectral parameter. Moreover, the adjoint problem of \eqref{4.5a} and \eqref{4.5b} reads
\begin{subequations}\label{4.6}
\begin{align}
\begin{pmatrix}
\widehat{\phi}_1 & \widehat{\phi}_2
\end{pmatrix}_x
&=-
\begin{pmatrix}
\widehat{\phi}_1 & \widehat{\phi}_2
\end{pmatrix}M,\label{4.6a}\\
\begin{pmatrix}
\widehat{\phi}_1 & \widehat{\phi}_2
\end{pmatrix}_t
&=-
\begin{pmatrix}
\widehat{\phi}_1 & \widehat{\phi}_2
\end{pmatrix}N. \label{4.6b}
\end{align}
\end{subequations}
\begin{Remark}
If $\begin{pmatrix} \phi_1&\phi_2 \end{pmatrix}^T$ is a solution to the linear problem \eqref{4.5a} and \eqref{4.5b}, then $\begin{pmatrix}\phi_2 & -\phi_1\end{pmatrix}$ is a solution to the adjoint problem \eqref{4.6a} and \eqref{4.6b}.
\end{Remark}

Let us first compute the gradient of spectral parameter $\eta$ with respect to $m$ and $n$, denoted by $\begin{pmatrix}\delta_m \eta & \delta_n \eta\end{pmatrix}^T$. From \eqref{4.5a}, the directional derivatives of $\phi_1$ and $\phi_2$ in the direction $m+\epsilon\Delta m$ are given by
\begin{equation}\label{4.7}
\begin{aligned}
\begin{pmatrix}
\phi_1^{\prime}[\Delta m]\\[0.4em]
\phi_2^{\prime}[\Delta m]
\end{pmatrix}_x
=&\begin{pmatrix}
- \frac{1}{2} &  \frac{1}{2}\eta m\\[0.4em]
- \frac{1}{2}\eta n &  \frac{1}{2}
\end{pmatrix}
\begin{pmatrix}
\phi_1^{\prime}[\Delta m]\\[0.4em]
\phi_2^{\prime}[\Delta m]
\end{pmatrix}\\
&+
\begin{pmatrix}
0 &  \frac{1}{2}\eta \Delta m+\frac{1}{2}m<\Delta m, \delta_m \eta>   \\[0.4em]
- \frac{1}{2}n<\Delta m, \delta_m \eta> &  0
\end{pmatrix}
\begin{pmatrix}
\phi_1\\
\phi_2
\end{pmatrix},
\end{aligned}
\end{equation}
where the pairing $\langle \Delta m, \delta_m \eta \rangle$ is defined as $\langle \Delta m, \delta_m \eta \rangle=\int \Delta m (\delta_m \eta)dx$. Left-multiplying both sides of \eqref{4.7} by $\begin{pmatrix}\widehat\phi_1 & \widehat\phi_2\end{pmatrix}$ and integrating over $x$, we get
\begin{equation}\label{4.8}
\begin{aligned}
\int \begin{pmatrix}\widehat\phi_1 & \widehat\phi_2\end{pmatrix}&\begin{pmatrix}
\phi_1^{\prime}[\Delta m]\\[0.4em]
\phi_2^{\prime}[\Delta m]
\end{pmatrix}_x dx
=\int \begin{pmatrix}\widehat\phi_1 & \widehat\phi_2\end{pmatrix}\begin{pmatrix}
- \frac{1}{2} &  \frac{1}{2}\eta m\\[0.4em]
- \frac{1}{2}\eta n &  \frac{1}{2}
\end{pmatrix}
\begin{pmatrix}
\phi_1^{\prime}[\Delta m]\\[0.4em]
\phi_2^{\prime}[\Delta m]
\end{pmatrix} dx \\
&+\int \begin{pmatrix}\widehat\phi_1 & \widehat\phi_2\end{pmatrix}\begin{pmatrix}
0 &  \frac{1}{2}\eta \Delta m+\frac{1}{2}m<\Delta m, \delta_m \eta>      \\[0.4em]
- \frac{1}{2}n<\Delta m, \delta_m \eta> &  0
\end{pmatrix}
\begin{pmatrix}
\phi_1\\
\phi_2
\end{pmatrix} dx.
\end{aligned}
\end{equation}
Integrating the left-hand side of \eqref{4.8} by parts under the assumption that boundary terms vanish, and using  \eqref{4.6a}, we find
\begin{equation}\label{4.9}
\int \begin{pmatrix}\widehat\phi_1 \quad \widehat\phi_2\end{pmatrix}\begin{pmatrix}
0 &  \frac{1}{2}\eta \Delta m+\frac{1}{2}m<\Delta m, \delta_m \eta>    \\[0.4em]
- \frac{1}{2}n<\Delta m, \delta_m \eta> &  0
\end{pmatrix}
\begin{pmatrix}
\phi_1\\
\phi_2
\end{pmatrix} dx=0,
\end{equation}
which is equivalent to
\begin{equation}\label{4.10}
\int \eta \Delta m \widehat{\phi}_1\phi_2 dx-<\Delta m,\delta_m \eta>\int n\phi_1\widehat{\phi}_2-m\widehat{\phi}_1\phi_2dx=0.
\end{equation}
Since $\Delta m$ is arbitrary, the variational derivative of $\eta$ with respect to $m$ is
\begin{equation}\label{4.11}
\delta_m \eta=\frac{\eta\widehat{\phi}_1\phi_2}{\int n\phi_1\widehat{\phi}_2-m\widehat{\phi}_1\phi_2dx}.
\end{equation}
Similarly, the variational derivative of $\eta$ with respect to $n$ is
\begin{equation}\label{4.12}
\delta_n \eta=-\frac{\eta\phi_1\widehat{\phi}_2}{\int n\phi_1\widehat{\phi}_2-m\widehat{\phi}_1\phi_2dx}.
\end{equation}
The presence of a common constant denominator in $\delta_m \eta$ and $\delta_n \eta$ is inessential for our purpose and is omitted. Hence, the gradient of the spectral parameter $\eta$ simplifies to
\begin{equation}\label{4.13}
\begin{pmatrix}
\delta_m \eta\\[0.4em]
\delta_n \eta
\end{pmatrix}
\propto
\begin{pmatrix}
\widehat{\phi}_1\phi_2\\[0.4em]
-\phi_1\widehat{\phi}_2
\end{pmatrix}.
\end{equation}

According to the general theory of Hamiltonian systems \cite{fuchssteiner,fokas1982}, applying Hamiltonian operator $\mathcal{D}_1$ to the gradient of spectral parameter $\eta$ leads to
\begin{equation}\label{4.14}
\begin{aligned}
\begin{pmatrix}
\Omega^m\\[0.4em]
\Omega^n
\end{pmatrix}
&=\mathcal{D}_1
\begin{pmatrix}
\widehat{\phi}_1\phi_2\\[0.4em]
-\phi_1\widehat{\phi}_2
\end{pmatrix}\Biggm|_{\eqref{4.5a}\eqref{4.6a}}  \\
&=
\begin{pmatrix}
\frac{1}{2}\eta (m_x-m)(\phi_1 \widehat{\phi}_1-\phi_2 \widehat{\phi}_2)+\frac{1}{2}\eta^2 m(n\phi_1 \widehat{\phi}_2+m\phi_2 \widehat{\phi}_1)\\[0.4em]
\frac{1}{2}\eta (n_x+n)(\phi_1 \widehat{\phi}_1-\phi_2 \widehat{\phi}_2)+\frac{1}{2}\eta^2 n(n\phi_1 \widehat{\phi}_2+m\phi_2 \widehat{\phi}_1)
\end{pmatrix}.
\end{aligned}
\end{equation}
It follows from the relations $m=u-u_{xx}$ and $n=v-v_{xx}$ that
\begin{equation}\label{4.15}
\Omega^m=(1-\partial_x^2)\phi_1\widehat{\phi}_2=\Omega^u-\Omega^u_{xx},\quad \Omega^n=(1-\partial_x^2)\widehat{\phi}_1\phi_2=\Omega^v-\Omega^v_{xx}.
\end{equation}
Therefore, we set
\begin{equation}\label{4.16}
\Omega^u=\phi_1\widehat{\phi}_2,\quad \Omega^v=\widehat{\phi}_1\phi_2.
\end{equation}
\begin{Proposition}
Let $\begin{pmatrix} \phi_1 &\phi_2\end{pmatrix}^T$ be determined by linear problem \eqref{4.5a} and \eqref{4.5b}, $\begin{pmatrix}\widehat{\phi}_1& \widehat{\phi}_2\end{pmatrix}$ be determined by adjoint problem \eqref{4.6a} and \eqref{4.6b}, then $(\Omega^u ,\Omega^ v, \Omega^m, \Omega^n)$ defined by \eqref{4.14} and \eqref{4.16} is a nonlocal symmetry of the two-component CH system with cubic nonlinearity \eqref{4.1}.
\end{Proposition}
Following Remark 4.1, the substitution $\begin{pmatrix}\widehat{\phi}_1& \widehat{\phi}_2\end{pmatrix}=\begin{pmatrix}\phi_2 &  -\phi_1\end{pmatrix}$ in \eqref{4.14} and \eqref{4.16} yields a reduced nonlocal symmetry of the system \eqref{4.1}.
\begin{Corollary}
Let $\begin{pmatrix} \phi_1 &\phi_2\end{pmatrix}^T$ be determined by \eqref{4.5a} and \eqref{4.5b}, then $(\omega^u ,\omega^ v, \omega^m, \omega^n)$ is a nonlocal symmetry of the two-component CH system with cubic nonlinearity \eqref{4.1}, where
\begin{equation}\label{4.17}
\begin{pmatrix}
\omega^u\\[0.4em]
\omega^v\\[0.4em]
\omega^m\\[0.4em]
\omega^n
\end{pmatrix}
=
\left.
\begin{pmatrix}
\Omega^u\\[0.4em]
\Omega^v\\[0.4em]
\Omega^m\\[0.4em]
\Omega^n
\end{pmatrix}
\right|_{\begin{pmatrix}\widehat{\phi}_1& \widehat{\phi}_2\end{pmatrix}=\begin{pmatrix}\phi_2 &  -\phi_1\end{pmatrix}}
=
\begin{pmatrix}
-\phi_1^2\\[0.4em]
\phi_2^2\\[0.4em]
\eta (m_x-m)\phi_1 \phi_2+\frac{1}{2}\eta^2 m(m\phi_2^2-n\phi_1^2)\\[0.4em]
\eta (n_x+n)\phi_1 \phi_2+\frac{1}{2}\eta^2 n(m\phi_2^2-n\phi_1^2)
\end{pmatrix}.
\end{equation}
\end{Corollary}

Suppose that linear problem \eqref{4.5a} and \eqref{4.5b} is invariant (up to the first degree of $\varepsilon$) under the infinitesimal transformation
\begin{equation}\label{4.18}
\begin{aligned}
u&\mapsto u+\varepsilon\omega^u,\quad &v&\mapsto v+\varepsilon\omega^v,\quad &m&\mapsto m+\varepsilon\omega^m,\\
n&\mapsto n+\varepsilon\omega^n,\quad &\phi_1&\mapsto \phi_1+\varepsilon\omega_1,\quad &\phi_2&\mapsto \phi_2+\varepsilon\omega_2,
\end{aligned}
\end{equation}
where $\omega^u$, $\omega^v$, $\omega^m$ and $\omega^n$ are given by \eqref{4.17}. Then $\omega_1$ and $\omega_2$ are determined by solving the linearized equations
\begin{subequations}\label{4.19}
\begin{align}
\begin{pmatrix}
\omega_1\\[0.4em]
\omega_2
\end{pmatrix}_x
&=\frac{dM[u+\varepsilon\omega^u,v+\varepsilon\omega^v,m+\varepsilon\omega^m,n+\varepsilon\omega^n;\eta]}{d\varepsilon}\biggm|_{\varepsilon=0}
\begin{pmatrix}
\phi_1\\[0.4em]
\phi_2
\end{pmatrix}
+M[u,v,m,n;\eta]
\begin{pmatrix}
\omega_1\\[0.4em]
\omega_2
\end{pmatrix},\label{4.19a}\\
\begin{pmatrix}
\omega_1\\[0.4em]
\omega_2
\end{pmatrix}_t
&=\frac{dN[u+\varepsilon\omega^u,v+\varepsilon\omega^v,m+\varepsilon\omega^m,n+\varepsilon\omega^n;\eta]}{d\varepsilon}\biggm|_{\varepsilon=0}
\begin{pmatrix}
\phi_1\\[0.4em]
\phi_2
\end{pmatrix}
+N[u,v,m,n;\eta]
\begin{pmatrix}
\omega_1\\[0.4em]
\omega_2
\end{pmatrix}.\label{4.19b}
\end{align}
\end{subequations}
To solve them, we introduce a new pseudo-potential $p$ satisfying
\begin{subequations}\label{4.20}
\begin{align}
p_x&=-\frac{1}{2}\eta^2 m\phi_2^2,\label{4.20a}\\
p_t&=-\frac{1}{\eta}\phi_1\phi_2+\frac{1}{2}(v+v_x)\phi_1^2-\frac{1}{4}\eta^2 (uv-u_x v_x)m\phi_2^2,\label{4.20b}
\end{align}
\end{subequations}
from which a solution to \eqref{4.19a} and \eqref{4.19b} is
\begin{equation}\label{4.21}
\begin{pmatrix}
\omega_1\\[0.4em]
\omega_2
\end{pmatrix}
=
\begin{pmatrix}
\phi_1p+\eta\phi_1\phi_{1x}\phi_2+\frac{1}{2}\eta\phi_1^2\phi_2\\[0.4em]
\phi_2p+\eta\phi_1\phi_2\phi_{2x}+\frac{1}{2}\eta\phi_1\phi_2^2
\end{pmatrix}.
\end{equation}
Furthermore, assuming the system \eqref{4.20a} and \eqref{4.20b} remains invariant under the infinitesimal transformation \eqref{4.18} along with $p\mapsto p+\varepsilon\omega^p$, we have
\begin{equation}\label{4.22}
\omega^p=p^2+\eta\phi_1\phi_2p_x.
\end{equation}
Consider an enlarged system consisting of \eqref{4.1}, \eqref{4.5a}, \eqref{4.5b}, \eqref{4.20a} and \eqref{4.20b}, the nonlocal symmetry $(\omega^u ,\omega^ v, \omega^m, \omega^n)$ in Corollary 4.3 can be prolonged to this system, which allows us to obtain a nonlocal symmetry of the enlarged system in terms of an evolutionary vector field
\begin{equation}\label{4.23}
\begin{aligned}
&-\phi_1^2\frac{\partial}{\partial u}+\phi_2^2\frac{\partial}{\partial v}+\left(\phi_1^2-\eta m\phi_1\phi_2\right)\frac{\partial}{\partial u_x}+\left(\phi_2^2-\eta n\phi_1\phi_2\right)\frac{\partial}{\partial v_x}+\left(p^2+\eta\phi_1\phi_2p_x\right)\frac{\partial}{\partial p}\\
&+\left(\eta (m_x-m)\phi_1 \phi_2+\frac{1}{2}\eta^2 m(m\phi_2^2-n\phi_1^2)\right)\frac{\partial}{\partial m}
+\left(\eta (n_x+n)\phi_1 \phi_2+\frac{1}{2}\eta^2 n(m\phi_2^2-n\phi_1^2)\right)\frac{\partial}{\partial n}\\
&+\left(\phi_1p+\eta\phi_1\phi_{1x}\phi_2+\frac{1}{2}\eta\phi_1^2\phi_2\right)\frac{\partial}{\partial \phi_1}
+\left(\phi_2p+\eta\phi_1\phi_2\phi_{2x}+\frac{1}{2}\eta\phi_1\phi_2^2\right)\frac{\partial}{\partial \phi_2},
\end{aligned}
\end{equation}
or equivalently in a non-evolutionary vector field form
\begin{equation}\label{4.24}
\begin{aligned}
\textbf{V}=&-\eta\phi_1\phi_2\frac{\partial}{\partial x}-\left(\phi_1^2+\eta\phi_1\phi_2 u_x\right)\frac{\partial}{\partial u}+\left(\phi_2^2-\eta\phi_1\phi_2 v_x\right)\frac{\partial}{\partial v}+\left(\phi_1^2-\eta u\phi_1\phi_2\right)\frac{\partial}{\partial u_x}\\
&+\left(\phi_2^2-\eta v\phi_1\phi_2\right)\frac{\partial}{\partial v_x}+p^2\frac{\partial}{\partial p}
+\left(-\eta m\phi_1 \phi_2+\frac{1}{2}\eta^2 m(m\phi_2^2-n\phi_1^2)\right)\frac{\partial}{\partial m}\\
&+\left(\eta n\phi_1 \phi_2+\frac{1}{2}\eta^2 n(m\phi_2^2-n\phi_1^2)\right)\frac{\partial}{\partial n}
+\left(\phi_1p+\frac{1}{2}\eta\phi_1^2\phi_2\right)\frac{\partial}{\partial \phi_1}
+\left(\phi_2p+\frac{1}{2}\eta\phi_1\phi_2^2\right)\frac{\partial}{\partial \phi_2}.
\end{aligned}
\end{equation}
The vector field $\textbf{V}$ acts as the generator of the one-parameter symmetry group for the enlarged system \eqref{4.1}, \eqref{4.5a}, \eqref{4.5b}, \eqref{4.20a} and \eqref{4.20b}. The symmetry transformation
\begin{equation}\label{4.25}
(\tilde{x}, \tilde{t}, \tilde{u}, \tilde{v}, \tilde{m}, \tilde{n}, \widetilde{\phi}_1, \widetilde{\phi}_2, \tilde{p})\equiv\exp(\varepsilon \textbf{V})(x, t, u, v, m, n, \phi_1, \phi_2, p),
\end{equation}
is explicitly formulated as
\begin{subequations}\label{4.26}
\begin{align}
\tilde{x}&=x+\ln \frac{1-\varepsilon p-\varepsilon\eta\phi_1\phi_2}{1-\varepsilon p},\quad \tilde t=t,\label{4.26a}\\
\tilde{u}&=(u+u_x)\frac{1-\varepsilon p-\varepsilon\eta\phi_1\phi_2}{2(1-\varepsilon p)}-(u_x -u)\frac{1-\varepsilon p}{2(1-\varepsilon p-\varepsilon\eta\phi_1\phi_2)}-\frac{\varepsilon \phi_1^2}{1-\varepsilon p-\varepsilon\eta\phi_1\phi_2},\label{4.26b}\\
\tilde{v}&=(v+v_x)\frac{1-\varepsilon p-\varepsilon\eta\phi_1\phi_2}{2(1-\varepsilon p)}-(v_x -v)\frac{1-\varepsilon p}{2(1-\varepsilon p-\varepsilon\eta\phi_1\phi_2)}+\frac{\varepsilon \phi_2^2}{1-\varepsilon p},\label{4.26c}\\
\tilde{m}&=\frac{2m(1-\varepsilon p-\varepsilon\eta\phi_1\phi_2)^2}{(1-\varepsilon p-\varepsilon\eta\phi_1\phi_2)[2(1-\varepsilon p)-\varepsilon \eta^2(m\phi_2^2-n\phi_1^2)]+\varepsilon^2 \eta^3 n\phi_1^3 \phi_2},\label{4.26d}\\
\tilde{n}&=\frac{2n(1-\varepsilon p)^2}{(1-\varepsilon p-\varepsilon\eta\phi_1\phi_2)[2(1-\varepsilon p)-\varepsilon \eta^2(m\phi_2^2-n\phi_1^2)]+\varepsilon^2 \eta^3 n\phi_1^3 \phi_2},\label{4.26e}\\
 \widetilde{\phi}_1&=\frac{\phi_1}{\sqrt{(1-\varepsilon p)(1-\varepsilon p-\varepsilon\eta\phi_1\phi_2)}},\quad  \widetilde{\phi}_2=\frac{\phi_2}{\sqrt{(1-\varepsilon p)(1-\varepsilon p-\varepsilon\eta\phi_1\phi_2)}},\quad \tilde{p}=\frac{p}{1-\varepsilon p}\label{4.26f}.
\end{align}
\end{subequations}
\begin{Proposition}
The enlarged system \eqref{4.1}, \eqref{4.5a}, \eqref{4.5b}, \eqref{4.20a} and \eqref{4.20b} is invariant under the finite symmstry transformation \eqref{4.25}. More precisely, if $(x, t, u, v, m, n, \phi_1, \phi_2, p)$ is a solution of this enlarged system, then $(\tilde{x}, \tilde{t}, \tilde{u}, \tilde{v}, \tilde{m}, \tilde{n}, \widetilde{\phi}_1, \widetilde{\phi}_2, \tilde{p})$ defined by \eqref{4.26a}-\eqref{4.26f} is also a solution.
\end{Proposition}

Using the finite symmetry transformation \eqref{4.25}, we now construct nontrivial solutions for the system \eqref{4.1}. Starting from a trivial solution $(u,v,m,v)=(u_0,1,u_0,1)$ of the system \eqref{4.1}, where $u_0$ is a constant satisfying
\begin{equation}\label{4.27}
1-\eta^2 u_0>0,
\end{equation}
the corresponding special solutions to \eqref{4.5a}, \eqref{4.5b}, \eqref{4.20a} and \eqref{4.20b} are taken as
\begin{equation}\label{4.28}
\phi_1=e^{\frac{kz}{2}}, \quad \phi_2=\frac{1+k}{\eta u_0}e^{\frac{kz}{2}},\quad p=-\frac{(1+k)^2}{2ku_0}e^{kz},
\end{equation}
with $k=\sqrt{1-\eta^2 u_0}$ and $z=x+\frac{(3-k^2)t}{2\eta^2}$. By substituting them into \eqref{4.26a}-\eqref{4.26f}, we find the following nontrivial solution of the system \eqref{4.1}
\begin{equation}\label{4.29}
\begin{aligned}
\tilde{x}&=x+\ln |1-k|-\ln |1+k\theta|,\quad &\tilde t&=t,\\
\tilde{u}&=\frac{[2-k^2(1+\theta^2)]u_0}{2(1+k)(1+k\theta)},\quad &\tilde{v}&=\frac{1+k(k+2\theta)+(1+k\theta)^2}{2(1-k)(1+k\theta)},\\
\tilde{m}&=\frac{2u_0(1-k)}{1-k^2+(1+k\theta)^2},\quad &\tilde{n}&=\frac{2(1+k\theta)^2}{(1-k)[1-k^2+(1+k\theta)^2]},
\end{aligned}
\end{equation}
where
\begin{equation}\label{4.30}
\theta \equiv
\begin{cases}
\tanh\left(\dfrac{k}{2}\left[x+\dfrac{(3-k^2)t}{2\eta^2}\right]+\ln\sqrt{\dfrac{\varepsilon(1-k^2)}{2ku_0}}\right), & \varepsilon > 0, \\[1.5em]
\coth\left(\dfrac{k}{2}\left[x+\dfrac{(3-k^2)t}{2\eta^2}\right]+\ln\sqrt{\dfrac{(-\varepsilon)(1-k^2)}{2ku_0}}\right), & \varepsilon < 0.
\end{cases}
\end{equation}

\section{Concluding and remarks}
\indent \indent In this paper, we study system of partial differential equations with the type
\begin{equation}\label{5.1}
\left\{
\begin{aligned}
u_{t} - u_{xxt} &= F(x, t, u, u_{x}, \dots, \partial ^{m} u/\partial _x^{m}, v, v_{x}, \dots, \partial ^{n} v/\partial _x^{n}), \\
v_{t} - v_{xxt} &= G(x, t, u, u_{x}, \dots, \partial ^{m} u/\partial _x^{m}, v, v_{x}, \dots, \partial ^{n} v/\partial _x^{n}).
\end{aligned}
\right.
\end{equation}
Under certain assumption on the coefficients of the connection 1-form associated with the surfaces, we provide a classification of system \eqref{5.1} that describes a pseudospherical or spherical surface. In particular, the results yield a classification for the following special third-order system
\begin{equation}\label{5.2}
\left\{
\begin{aligned}
u_{t} - u_{xxt} &= A_1(u, u_{x},v,v_{x})u_{xxx}+B_1(u,u_{x},u_{xx},v,v_{x},v_{xx}), \\
v_{t} - v_{xxt} &= A_2(u, u_{x},v,v_{x})v_{xxx}+B_2(u,u_{x},u_{xx},v,v_{x},v_{xx}),,
\end{aligned}
\right.
\end{equation}
As examples, we show that series of systems belong to such class, such as the Song-Qu-Qiao system, the two-component CH system with cubic nonlinearity and the modified CH-type system.  For the two-component CH system with cubic nonlinearity \eqref{4.1}, we construct nonlocal symmetries from gradients of spectral parameter. By introducing an appropriate pseudo-potential, we prolong the reduced nonlocal symmetry to an enlarged system and thereby derive the corresponding finite symmetry transformation. On this basis, we calculate a nontrivial solution for the system \eqref{4.1}.

Systems of CH-type have natural geometric correspondence, such as the multi-component CH system \cite{kang2021}. It is worthwhile to see whether the argument used in this paper can be applied to a more general cases than \eqref{5.1}. Therefore, we can find  more CH-type systems which can be connected to pseudospherical or spherical surfaces. Meanwhile, a separate challenge concerns the Song-Qu-Qiao system, for which the nonlocal symmetry construction presented here is not applicable. This raises the natural question of how to systematically derive nonlocal symmetries for the Song-Qu-Qiao system.

\section*{Acknowledgements} Guo's research is supported by Northwest University Graduate Research and Innovation Program CX2024133. Kang's research is supported by NSFC (Grant No. 12371252) and Basic Science Program of Shaanxi Province (Grant No. 2019JC-28).


\end{document}